\documentclass[5p, authoryear]{elsarticle}
\pdfoutput=1

\usepackage{amsmath}
\usepackage[svgnames]{xcolor}

\newcommand{\tc}[1]{\textcolor{Black}{#1}}
\usepackage{algorithmic}
\usepackage{algorithm}

\usepackage{multicol, caption}

\usepackage{lipsum}
\usepackage{float}

\begin{document}

NOTICE: this is the author's version of a work that was accepted for publication in Geoderma. Changes resulting from the publishing process, such as peer review, editing, corrections, structural formatting, and other quality control mechanisms may not be reflected in this document. 
Changes may have been made to this work since it was submitted for publication. A definitive version was subsequently published in Geoderma, Volumes 223-225, July 2014, Pages 97-107  doi:10.1016/j.geoderma.2014.01.005

\begin{frontmatter}
\title{Evaluation of modelling approaches for predicting the spatial distribution of soil organic carbon stocks at the national scale}

\author[1]{M.~P.~Martin}
\author[2]{T.G.~Orton}
\author[1]{E.~Lacarce}
\author[3]{J.~Meersmans}
\author[1]{N.P.A.~Saby}
\author[1]{J.B.~Paroissien}
\author[1]{C.~Jolivet}
\author[1]{L.~Boulonne}
\author[1]{D.~Arrouays}

M.P. Martin
T.G. Orton
E. Lacarce
J. Meersmans
N.P.A. Saby
J.B. Paroissien
C. Jolivet
L. Boulonne
D. Arrouays

\address[1]{INRA, US1106 Unit\'e Infosol, F-45000 Orl\'eans, France}
\address[2]{Faculty of Agriculture and Environment, The University of Sydney, 1
Central Avenue, Australia Technology Park, Eveleigh, NSW 2015, Australia}
\address[3]{Department of Geography, College of Life and Environmental Sciences, University of
Exeter. Amory building - room 431, Rennes Drive EX4 4RJ Exeter - UK}


\begin{abstract}
Soil organic carbon (SOC) plays a major role in the global carbon budget. It can act as a source or a sink of
atmospheric carbon, thereby possibly influencing the course of climate change.
Improving the tools that model  the spatial distributions of SOC stocks at national scales is a priority,
both for monitoring changes in SOC and as an input for global carbon cycles studies.
In this paper, we compare and evaluate two recent and promising modelling approaches. First, we considered 
several increasingly complex boosted regression trees (BRT), a convenient and efficient multiple
regression model from the statistical learning field. Further, we considered a robust geostatistical
approach coupled to the BRT models. Testing the different approaches was performed on the dataset
from the French Soil Monitoring Network, with a consistent cross-validation procedure. 
We showed that when a limited number of predictors were included in the BRT model, 
 the standalone BRT predictions were significantly improved by robust geostatistical modelling of the
 residuals. However, when data for several SOC drivers were included, the standalone BRT
 model predictions were not significantly improved by geostatistical modelling. Therefore, in this
 latter situation, the BRT predictions might be considered adequate without the need for geostatistical
 modelling, provided that i) care is exercised in model fitting and validating, and ii) the dataset
 does not allow for modelling of local spatial autocorrelations, as is the case for many national
 systematic sampling schemes.

\end{abstract}
\end{frontmatter}

\section{Introduction}
Soils are the second biggest carbon pool of the planet, containing about 1500\,PgC 
\citep{Batjes1996,Eswaran1993,POST1982}. As such, their behaviour as a greenhouse gas source and sink needs
to be quantified, when facing climate change induced by increasing atmospheric greenhouse gases concentrations 
\citep{Batjes1996,Lal2004}.
Quantifying temporal changes of this pool requires estimating its spatial distribution at different dates
and at various scales, with the national scale being of particular importance for international
negotiations. 
The reliability of such estimates depends upon suitable data in terms of organic carbon content and soil bulk 
density and on the methods used to upscale point data to comprehensive spatial estimates. 
These estimates may also be used for defining the baseline state for soil organic
carbon (SOC) change simulations 
\citep{van_Wesemael2010}, or setting some of 
the parameters for models of SOC dynamics \citep{Tornquist2009}.


Interestingly, there is quite a diversity regarding the nature of the models used for upscaling SOC point
measurements to the national level. The validity of each method depends on the datasets and on the scale
(defined by its grain or precision and extent, \citealp{Turner1989}).
The mapping approaches range from \mbox{simple} statistics or pedotransfer rules, relating SOC
contents or stocks to soil type \citep{Yu2007} or soil type and land use
\citep{Tomlinson2006, 48}, to multivariate regression models (\citealp[with
multiple linear models]{Meersmans2008} and \citealp[with generalized linear
models]{Yang2008} or \citealp[with mixed models]{Suuster2012}). Recent studies have used techniques adapted from the data
mining and machine learning literature, with piecewise linear tree models
\citep{Bui2009} or multiple regression trees for regional studies
\citep{Grimm2008, Lo_Seen2010, Suuster2012}. 
Among the studies considering small extent (<50 km$^2$),
many have considered the use of geostatistics, some including SOC predictors $via$ cokriging (CK) or regression 
kriging (RK) \citep{Mabit2010, Don2007, Rossi2009, Wang2009, Spielvogel2009}. 
As the extent increases, the use of geostatistics becomes less common and 
despite the spatial dimension of such studies, few geostatistical approaches for SOC mapping have been proposed
for use at the national scale (but see \citealp{CHAPLOT2009, Kerry2012, Rawlins2009}).
\\


SOC mapping for France has been performed, during the last decade, by using class specific SOC means \citep{48}
or regression models \citep{bg-8-1053-2011, Meersmans2012}. The most recently proposed models are still 
not able to fully satisfactorily predict SOC stocks or contents on independent
locations :
R$^2$ reached 0.50 and 0.49 and root mean squared prediction errors (RMSPE)
 2.27~kg/m$^2$ and 1.45\%, for \citet{bg-8-1053-2011} on SOC stocks and for \citet{Meersmans2012}
on SOC contents, respectively.
\citet{bg-8-1053-2011} obtained unbiased predictions (the bias was estimated to be -0.002~kg/m$^2$
by cross-validation), which might ensure unbiased mapping of the stock at
the national level. Nevertheless, these R$^2$ and RMSPE results showed that there
is potentially room for improvement, especially if one is willing to use such models for regional
assessments. Adding spatial autocorrelation terms in these models 
might be a way to improve their performance.

Recently, new approaches have been proposed for coupling regression models, relating
environmental factors to the studied property,
with geostatistical models, representing the spatial autocorrelation among the
observations \citep[see for example][]{Marchant2010}. Such 
methods were also designed to handle local anomalies (\emph{i.e.} outliers).
Nevertheless, these methods do not currently include some features that other statistical 
models, such as boosted regression trees (BRT) used by \citet{bg-8-1053-2011},
have (\emph{i.e.} handling nonlinear relationships between
qualitative and quantitative predictors and the independent variable, nonlinear
interactions between the predictors, in an automated manner).
Both approaches share the robustness to the presence of outliers in the dataset. As they are tackling different problems,
the spatial autocorrelation for the geostatistical approaches, and the modelling of the complex interactions 
between SOC stocks and their drivers for the regression methods, both might be considered as complementary.

The aim of this paper is to combine these recent robust
geostatistical approaches with the BRT models currently applied to map SOC
stocks at the national scale  for France. 
We apply the methods to a dataset of 2166 paired observations of SOC and bulk
 densities from the French soil quality monitoring network (RMQS). We use this
 study to assess the modelling methods to determine i) how useful it is to
 combine BRT and geostatistical modelling, and ii) if any advantages are dependent
 on the number of ancillary variables included as predictors in the BRT models. 
The aim is not specifically to study the relative importance of SOC stocks drivers for France
\citep[which has been done recently][]{bg-8-1053-2011, Meersmans2012}, nor to produce
a new map of SOC stocks in France.

\section{Materials and methods}
\subsection{Data}\label{ssv}

\begin{figure}
  \includegraphics[width=80mm]{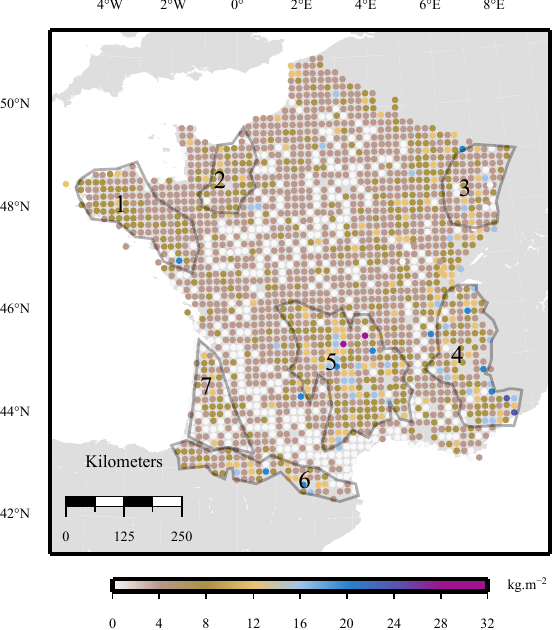}
  \captionof{figure}{SOC stocks (0-30cm) values on the French monitoring network, which were used in the present study.
Areas from 1 to 7 represent various different areas that are mentioned later in the text. 1:~south-west Brittany.
2:~part of Basse Normandie. 3:~Alsace and part of Lorraine. 4:~part of French Alps. 5:~Massif Central. 6:~French
Pyrenean mountain range. 7:~ part of Aquitaine.}
\label{mape}
\end{figure}

Soil Organic Carbon Stocks were computed for 2166 sites from the
French soil quality monitoring network (RMQS) (Fig.~\ref{mape}).
The network is based on a 16\,km\,$\times$\,16\,km square grid.
The sampling sites are located at the center of each grid cell, except when 
settling a homogeneous 20\,m\,$\times$\,20\,m sampling area is not possible at this specific location
 (because of the soils being sealed or strongly disturbed by anthropogenic
activities, for instance). 
In that case, another site is selected within 1\,km from the
center of the cell depending on soil availability for sampling \citep[for more
information, see][]{Arrouays2002}.
 Some of the 2166 sites of our dataset 
were actually replicates of the regular cells sites : some cells had two sites
located in them, one close to the center of the cell as described above, and another
 one located at another position within the cell.

\tc{At each site, 25 individual core samples were taken from the 
(0--30\,cm) and the subsoil (30--50\,cm) using a hand auger according to an unaligned sampling 
design within a 20\,m\,$\times$\,20\,m area.
Individual samples were mixed to obtain a composite sample for each soil layer.
In addition to the composite sampling, a soil pit was dug 5\,m from the south border
of the 20\,m\,$\times$\,20\,m area, from which 6 bulk density measurements were done, as described
previously \citep{Martin2009}. }
 From these data, SOC stocks (kg/m$^2$) were 
computed for the 0--30\,cm soil layer~:

\begin{equation}
\mathit{\rm SOC {\rm stocks}}_{30\;{\rm cm}}=\sum
_{i=1}^{n}{p_{i}{\rm BD}_{i}{\rm SOC}_{i}(1-{rf}_{i})}
\end{equation}
where $n$ is the number of soil horizon present in the 0--30\,cm layer,
BD$_i$, \textit{rf}$_i$ and SOC$_i$ the bulk
density, percentage of rock fragments (relative to the mass of soil) and the
SOC concentration (percent) in these horizons, and $p$$_i$ the
width of the horizons to take into account to reach the 30\,cm. The horizons 
considered for such an analysis did not include the organic horizons (such as OH or OL).


The SOC stocks, the dependent variable, is the only variable which was observed at site level.
All other variables, the covariates (or ancillary variables) were depicted using available maps
covering the French territory. This allowed us to consider models for mapping SOC
distributions at the national scale, which relies on the exhaustiveness of
the ancillary information.
These ancillary maps were thus sampled at RMQS sites locations in order to estimate
climatic, pedological, land use and management related, and biological variables.

The map of pH was derived from two sources. For the forest soils, the forest soils surface pH map
\citep{pHMapForest2008} was used. For the other soils, the median pH per district from the national 
soil testing database  was used \citep[BDAT, ][]{Lemercier2006}.
Land use was estimated from from Corine Land Cover 2006 database 
and further reclassed into an  adapted IPCC land use classification (various
crops, permanent grasslands, woodlands, orchards and shrubby perennial crops,
wetlands, others  and vineyards) \citep{UE-SOeS2006}.
Clay content was estimated from the 1:1 000 000 scale European
Soil Geographical Database \citep{King1995}. As each polygon (or soil unit) from the
1:1~000~000 scale European Soil Geographical map is linked to possibly several soil types (hence clay levels),
we used in the models the clay levels of the 3 most important (in terms of surface) 
soil types within each soil unit associated to each RMQS site, namely $clay1$, 
$clay2$ and $clay3$, ranked according to the percentage of their occurrence. Surface percentages 
of these soil types were also included as predictors within the models ($pc1$, $pc2$
and $pc3$). For 
instance, let us consider a given RMQS site $i$ belonging to soil unit $j$ of the soil map.
The soil unit $j$ may have two soil types associated to it ($st1$ and $st2$) with the 
occurring probabilities of 70\% and 30\% and clay levels of 45\% and 35\% . For this site $i$, the values of the 
$clay1$, $clay2$ and $clay3$ variables would be 45\%, 35\% and $NA$ (not available) respectively and
for $pc1$, $pc2$ and $pc3$, 70\% and 30\%  and $NA$ respectively.
Organic matter additions (\emph{oma}), such as slurry and farmyard manure were estimated.
We used manure application and animal excrement production departmental statistics \citep{ADEME2007}.
These statistics were combined with dry matter C concentration values,
\citep[37.7~\% for farm yard manure and 36.6~\% for slurry,]{Meersmans2012}.

Climatic data were monthly precipitation (mm\,month$^{-2}$), potential evapotranspiration (PET,
mm\,month$^{-2}$), and temperature ($^\circ$C) at each node of a
8\,$\times$\,8\,km$^2$ grid, averaged for the 1992--2004
period. This climatic map was obtained by interpolating observational data
using the SAFRAN model \citep{Quintana-Segui2008}. Again, for the 
modelling study presented here, climatic variables were estimated at 
each RMQS site by performing a spatial join between the RMQS grid and the climatic map.

Agro-pedo-climatic variables  were also derived from the
primary soil, climate and land use data estimated at each RMQS site: 
we used the ($a$) temperature and ($b$) soil  moisture mineralization modifiers,
as modelled in the RothC model \citep{coleman1997, bg-8-1053-2011}. The $b$ variable 
was calculated by combining, for each RMQS site, rainfall and PET data
obtained from the climatic grid, with site observation of land use and clay
content. Since three possible clay contents were estimated for each site,
the three corresponding estimates of the $b$ variables were also included, when relevant,
in our BRT models.
Lastly, the Moderate Resolution Imaging Spectroradiometer Net Primary Productivity
(MODIS NPP, gC\,m$^2$\,yr) was used to get NPP estimates
 at each of the RMQS sites, as in \citep{bg-8-1053-2011}.

The GIS processing was carried out using the GRASS GIS (GRASS Development Team, 2012)
and further mapping was carried out using Generic Mapping Tools software
\citep{Wessel1991}.\\

\subsection{Statistical modelling}
\subsubsection{Boosted Regression Trees (BRT) modelling}
Boosted regression trees belong to the Gradient Boosting Modelling (GBM) family.
The objective is to estimate the function $F$ that maps the values of a set of predictor
variables $x = \{x1, .., xp\}$ onto the values of the output variable $y$,
by minimizing a specified loss function $L$. 
This $L$ function is applied at each iteration in order to fit so-called 
base learners.
The final prediction of the BRT model is a linear combination of each
base learner prediction. The constant weight associated with these base learner
predictions is called the learning rate and is one of the important parameters
of this boosting algorithm \citep{Freund96experimentswith}. This 
kind of algorithm is also referred to as a forward ``stagewise'' procedure.
The base learners of BRT are classification and regression trees \citep{Brei1984}. 
Furthermore, BRT uses a specialized form (for regression trees) of Stochastic Gradient
Boosting \citep{Friedman2001}.
The stochastic characteristic of the algorithm relies on the fact that only a subset 
of the dataset is used for fitting the base learner on a given iteration. The subset 
is produced in each iteration using a uniform random draw without replacement. 
Besides the learning rate, other parameters are important when applying this
kind of model. Two of them determines the characteristics of each base learner :
the tree size (which gives the size of individual regression trees) and the minimum number of observations in the
terminal leaves of the trees. Several options are available for deciding
when to stop adding base leaners to the model. One of them, 
based on an internal cross-validation, was shown to be the most efficient one
\citep{Ridgeway2006} for avoiding overfitting and was used for the present study.
BRT was shown to have improved accuracy compared with simple regression trees,
thanks to its stochastic gradient boosting procedure aimed at minimizing the risk of overfitting
 and improving its predictive power \citep{Lawrence2004}. It can handle non-linear
interactions among predictors and the dependent variable, quantitative and qualitative
predictors and missing data.
Lastly, several tools are available for interpreting the behavior and characteristic
of the resulting BRT models, such as  the variable importance index
 for assessing the contribution of the predictors and the partial dependence plots
for assessing the relationships between predictors and the predicted variable  \citep{bg-8-1053-2011}.

A thorough description of
the method is given in \citet{Friedman2001} and a practical guide for using
it in \citet{Elith2008}.
The BRT models were fitted and used for prediction using the ``gbm'' R  \citep{RCore}
package \citep{Ridgeway2006}.

\subsubsection{Three BRT Models for SOC stocks}\label{mods}
Three models for predicting SOC stocks in the 0--30\,cm layer were tested. The models,
which we refer to as the LU, L and F models, have increasing levels of complexity
 (see below for their full description of these models). These three models
were chosen as they represent cases where either very little or a lot of information 
on ancillary variables is known on sites where SOC stocks are to be predicted.
Additionally, the first model (LU), with two covariates (landuse and clay
content) commonly
used for predicting SOC within the geostatistical framework. The second one (L, see
below the full description) is indeed the $Extra$ model presented in \cite{bg-8-1053-2011}. The use of the most complex model (F) enabled us to 
include all the ancillary data available for France at the national level at the time of
the present study.

The predictors used for each model were:
\begin{itemize}
\item \emph{LU}: \textit{lu\_ipcc} (land use classification adapted from the IPCC
guidelines, 2006), $clay1$.
\item \emph{L}:
\textit{lu\_ipcc}, $clay1$, $clay2$ and $clay3$, $pc1$, $pc2$
and $pc3$, the clay and corresponding probability of occurrences at each RMQS site,
(\textit{pet}, mm\,month$^{-2}$), monthly precipitation (\textit{rain}, mm\,month$^{-2}$), temperature
(\textit{temp}, $^\circ$C), the two RothC mineralization
modifiers, \textit{a} and $b1$, $b2$ and $b3$ and the net primary productivity \textit{npp}
(gC\,m\textsuperscript{$-$2}\,yr\textsuperscript{$-$1}).
\item \emph{F}: same predictors as the model L with the addition of \textit{pH},  \textit{oma} (\emph{i.e.} 
organic matter addition, slurry and farmyard manure) and \textit{pm}, the parent material.
\end{itemize}

The ``gbm'' R package requires the specification of several important parameters : 
the tree size, the learning rate, the minimum number
of observations in the terminal leaves of the trees and the bag fraction.
For our three models, the values for these parameters were set to $(12, 0.01, 3, 0.7)$. These values were 
chosen according to recommendations found in the literature \citep{Elith2008, Ridgeway2006}.

\subsubsection{Geostatistical models}\label{geost}
We further investigated whether a robust geostatistical method, similar to the one presented by \cite{saby2011}, 
could be used to represent errors and improve predictions from each of the BRT models.
In their work, Saby et al (2011) divided the spatial variation of a soil property into
 fixed and random effects. The fixed effects were a different constant mean soil property
for each of 12 parent material classes and the random effects described the spatially
correlated residual soil property variation. In the present work, 
each of the three BRT models was used alone as presented in the previous section and
as a fixed effect within a robust geostatistical method. This combination of
a BRT and geostatistical model can be summarised as:

\begin{equation}
\label{eq1}
\mathbf{Z}=H(\mathbf{X})+\mathbf{u} 
\end{equation}

where $\mathbf{Z}=ln(\mathbf{Y})$ with $\mathbf{Y}$ being a length n vector of observations of the SOC stocks,
 $\mathbf{X}$ the matrix (n~x~q) containing values of the covariates (or predictors) at each
observation site and the $H$ function representing the boosted regression tree model
(fitted to the log-transformed data). \
We note that the log-transform was necessary for the geostatistical approach due to
skewness of the observed SOC stocks distribution.
Thus the vector $\mathbf{u}$ of length n contains the residuals of the BRT model predictions of
the log-transformed data, compared to the log transformed response variable.
In the conventional geostatistical approach, these residuals are assumed to be a realization
of a second order stationary random process \citep{Webster2007}.
We applied a robust geostatistical approach, in which the spatial correlation of residuals was modelled using
a Mat\`{e}rn equation based on the Dowd robust estimator of the experimental variogram \citep{Dowd1984}.
	Moreover, outlying observations were identified  and Winsorized using the algorithm
 proposed by \cite{Hawkins1984}. 
Winsorizing is a method by which extreme values in the statistical data are limited
 to reduce the effect of possibly spurious outliers.
Note that Winsorizing is not equivalent to simply excluding data. 
Rather,  in a Winsorized procedure, the extreme values are replaced by a certain 
value predicted by a statistical model. 
This algorithm provided for each observed residual $u_i$ an interval $[U_i^-, U_i^+]$.
$u_i$ is then
identified as being an outlier when $u_i \notin [U_i^-, U_i^+]$ and its value is replaced
by the closest limit of the interval. As in \cite{Lacarce2012}, 
observations were confirmed as being outliers, and transformed, conditionally on a measurement
error of SOC stocks of $\epsilon Y$, with $\epsilon=0.112$, recently estimated  for the RMQS
dataset (unpublished data) :

\begin{equation}
\label{eqWin}
 u^*_i =
  \begin{cases}
   u_i^- & \text{if } ln(Y_i(1+\epsilon)) - H(\mathbf{X}_i) < U_i^- \\
   u_i^+ & \text{if } ln(Y_i(1-\epsilon)) - H(\mathbf{X}_i) > U_i^+ \\
   u_i & otherwise 
  \end{cases}
\end{equation}

where ui* represents the resulting Winsorized data. 
One should note that the geostatistical modelling is performed on the log scale, but
the measurement error is valid on the original scale, hence the terms 
on the left-hand side of the
inequalities in equation \ref{eqWin}. These inequalities mean that the observed
residuals may exceed the $[U_i^-, U_i^+]$ interval limits, but not by more than the possible
measurement error on the observed values. If they do, they are Winsorized to $U_i^-$
or $U_i^+$ depending on the case. 
The $U_i^-$ and $U_i^+$ values are defined so that the validity of the spatial term
$\mathbf{u}$ in equation \ref{eq1}
is verified, which, without Winsorizing, was rarely the case in previous
studies \citep{Marchant2010, Lacarce2012}. This check is performed using a
leave-one-out cross validation (LOOCV). 
When the covariance model is a valid representation of the spatial variation of the property (in our case the
residuals), the distribution of the squared standardized prediction errors (noted $\theta$) derived
from the cross validation will be a $\chi^{2}$ with mean $\overline{\theta}=1$ and median
$\breve{\theta}=0.455$ \citep{Marchant2010} for which confidence intervals may be
determined. This LOOCV procedure aims solely at checking the validity 
of the geostatistical model and should not be mistaken with the global validation
framework presented in the next section, aimed at estimating the predictive
performance of the models (BRT models and their spatial counterparts).


As a result the variation of the soil property is decomposed in a threefold
model described by \citet{Marchant2010}: 1) variation modelled by the BRT models, 2) spatially correlated
variation represented by the random effect of the residuals of the BRT models and estimated by variograms
using Dowd's estimator to which Mat\`{e}rn equations were fitted,
3) variation due to circumscribed anomalies.
Once the BRT and geostatistical models were fitted, the property was predicted at each unsampled 
(\textit{i.e} not used for fitting the models) location of the dataset by lognormal
ordinary kriging.
This method consists of predicting the residual for the log-transformed variable by ordinary kriging based on 
Winsorized data $\mathbf{u}^*$ (equation \ref{eqWin}),
and back-transforming the predicted value to the original SOC stocks scale through:
\begin{equation}
\label{eqlok}
\hat{Y}(\boldsymbol{x}_i)=exp(H(\mathbf{X}_i)+ \hat{u}_i + var[\hat{u}_i]/2 - \psi(\boldsymbol{x}_i))
\end{equation}

where $\hat{u}_i$ is the ordinary kriging prediction of u at a given prediction location $\boldsymbol{x}_i$,
$var[\hat{u}_i]$ is the associated kriging variance and $\psi$ the Lagrange multiplier;
both  the kriging variance and Lagrange multiplier are needed to yield unbiased estimates in case of lognormal
ordinary kriging \citep{Webster2007}.


%

\subsubsection{Validation procedure}
We thus considered six models: three models without a spatial term (the LU, L and F BRT models) and 
what is hereafter referred to as their spatial counterparts
(the LU$_g$, L$_g$ and F$_g$ models), i.e. the same three models with an additional spatial term (Eq. \ref{eq1}). 
These six models were 
validated using cross-validation.
This validation procedure involves validation against independent data and enables
estimation the predictive power 
of the proposed models.

Comparison between observed and predicted values of
SOC stocks was carried out on the original scale using several complementary
indices, as is
commonly suggested \citep{31}: the mean
prediction error (MPE, kg/m$2$), the root mean square prediction
error (RMSPE, kg/m$2$) and the
coefficient of determination ($R^{2}$) measuring
the strength of the linear relationship between predicted and observed
values. Additionally, the ratio of performance to inter-quartile distance,
RPIQ \tc{\citep{Bellon-Maurel2010}} was estimated as 
\begin{equation}
\label{eq2}
RPIQ = \frac{IQ_y}{RMSPE}
\end{equation}
where IQ$_y$ is the inter-quartile distance, calculated on observed SOC values from
the whole dataset. RPIQ index accounts much better for the spread of the population
than indexes such as RPD \citep{Bellon-Maurel2010} and was used for comparing the prediction accuracy between the six different models. 
Median prediction error and root of median of squared prediction errors 
were also calculated (hereafter named MedPE and RMedSPE respectively). 
These additions to MPE and RMSPE respectively, provide a more complete picture of the
errors in case of a skewed error distribution.\\

The validation procedure was done using a Monte Carlo 
10-fold cross-validation \citep{XU2001}, enabling us to perform what will be referred to in
the following as external validation. It was preferred to simple data-splitting
because the estimate of a model's performance then does not rely on the choice 
of a single sub-sample. We preferred a k-fold procedure 
instead of a leave-one-out cross-validation as leave-one-out cross-validation results 
in a high variance of the estimate of the prediction error \citep{hastie2001}. Each step of the 
cross-validation procedure can be summarized as shown in algorithm \ref{alg1} and 
was repeated 200 times for each model.


%

\begin{algorithm}
\caption{cross-validation repetition:}
\label{alg1}
\algsetup{indent=2em}
\begin{algorithmic}[1]
\STATE Split the dataset into Learning $(\mathbf{X}, \mathbf{Y})_{L}$ and Validation $(\mathbf{X}, \mathbf{Y})_{V}$
\STATE Compute $\mathbf{Z}_L = ln(\mathbf{Y}_L)$
\STATE Fit the $H$ BRT model and estimate $\hat{\mathbf{Z}}_{L}=H(\mathbf{X}_L)$
\STATE Fit a variogram on $\mathbf{u}_{L}=\mathbf{Z}_L - \hat{\mathbf{Z}}_L$
\IF {$\overline{\theta}$ and $\breve{\theta}$ are not valid} 
  \STATE Winsorize the dataset until valid $\overline{\theta}$ and $\breve{\theta}$ are obtained
\ENDIF
\STATE Estimate $\hat{\mathbf{u}}_{V}$ by ordinary kriging at $\mathbf{Z}_V$ locations using the fitted variogram and
the Winsorized residuals $\mathbf{u}^*_{L}$
\STATE Calculate the lognormal kriging estimate, $\hat{\mathbf{Y}}^{g}_{V}$ using equ.\ref{eqlok}
\STATE Calculate $\hat{\mathbf{Y}}_{V}=\exp(H(\mathbf{X}_V))$
\STATE Compute PERF on $(\mathbf{Y}_{V}, \hat{\mathbf{Y}}_{V})$ 
\STATE Compute PERF on $(\mathbf{Y}_{V}, \hat{\mathbf{Y}}^{g}_{V})$ 
\end{algorithmic}
\end{algorithm}

Steps 4 to 9 are performed as detailed in section \ref{geost}. More specifically, the spatial component
of the spatial models is validated at step 6 as presented in 
section \ref{geost} in order to make sure that these were valid representations of 
the residuals of the BRT models. 
This check was performed for each geostatistical model fitted during each repetition of the
cross-validation procedure.

$\hat{\mathbf{Y}}$ represents the prediction provided by one given BRT
 model and $\hat{\mathbf{Y}}^{g}$ 
the prediction provided by its spatial counterpart model (the BRT and the 
geostatistical model).
$PERF$ indicates the computation of the performance metrics (R$^2$, RMSPE, RPIQ, MPE,
 MedPE, RMedSPE).
We should note that the last step of the algorithm represents a true external validation of the spatial
model because the model fitting  is performed while masking the observations $Y_{V}$ used for
validation, both during the variogram fitting (step 4) and the kriging procedure
(step 8). 
A similar procedure has recently been used and advocated by \citet{Goov2010}, 
using leave-one-out cross-validation.
It should be distinguished from other approaches where cross-validation embeds only
the kriging, and not the fitting of variograms parameters (\textit{e.g.} 
\citealp{Wu2009, Mabit2010, Xie2011}). In these cases, observations used 
for validation have already been used for fitting the variogram and the resulting model is not independent
from these observations.
We tested for differences between the performances of the six models in terms of each performance metric.
 The distributions of a performance metric were compared  using a  t-test with a Bonferroni
adjustment. In the following, we use the terms MPE, RPIQ, RMSPE, R$^2$, MedPE and RMedSPE names
to refer to their mean value over the 200 repetitions of the cross-validation.
The algorithm procedure was programmed with R software using functionalities of geoR and sp packages 
\citep{Ribeiro2001, Bivand2008}. \\

\begin{figure}
\begin{center}
\includegraphics[width=85mm]{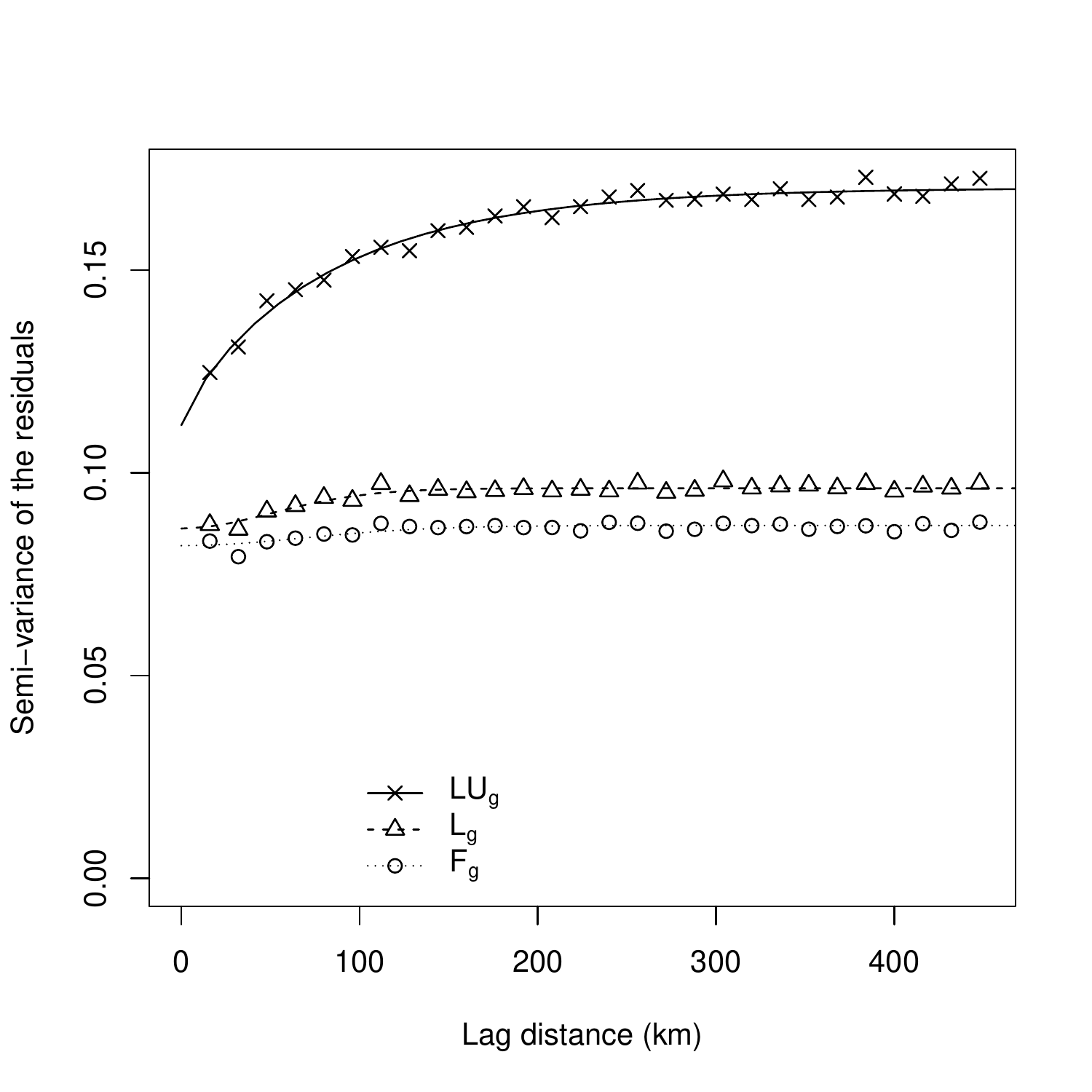}
\end{center}
\caption{Dowd variograms of the residuals of the LU, L and F models to estimate random effects
of the LU$_g$, L$_g$ and F$_g$ models. Residuals where calculated as the difference between 
the log transformed response variable and the BRT model predictions. The variograms
were obtained by fitting the Mat\`ern models on the full dataset.}
\label{variograms}
\end{figure}

\section{Results}
\subsection{Variogram fitting on BRT residuals}
The degree of spatial correlation of residuals from BRT models depended on the
complexity of BRT models  (Fig.\ref{variograms})~:
the residuals resulting from the LU$_g$ model were spatially structured with a spatial dependence (defined as 
partial sill/(nugget + partial sill), \citealp{Lark2004})
of 0.34. 
Contrary to the residuals of the LU$_g$ model, the residuals of the more complex BRT models exhibited 
very limited spatial structure (Fig.\ref{variograms}). 
For the F$_g$ model, the residuals had a spatial dependence of 
0.057
and for the L$_g$ model of 
0.1.
Fig.\ref{variograms} indicated that from the simplest model (LU$_g$) to the most complex one (F$_g$),
the part of the spatial variability not accounted for by the deterministic spatial trend decreased.

For the three models, Winsorizing was needed in order to produce valid models regarding the assumption
on the modelled variable. The $\overline{\theta}_w$ and $\breve{\theta}_w$ values obtained 
after Winsorizing belonged to the confidence interval estimated for
each model. The percentage of outliers ranged from 1.9\% to
2.8\%. 
These sites present extremely low or high SOC stocks that cannot be modelled by the spatial term and the BRT
models only.  
The number of such locations was halved between the LUg and the Fg models (Table \ref{varioFit})
as this latter model, the most complex one, was more able to model these extreme values.
For this model, outliers appeared to be evenly distributed over the studied area (Fig.\ref{residMaps}h).

\subsection{Cross-validation analysis \& performance of the proposed models}
Cross-validation yielded valid spatial models for 100\% of the cross-validation repetitions. 
Fig. \ref{CVPlot} shows that the F and L models and their spatial counterparts performed 
globally similarly to other and differently
from the LU and LU$_g$ model. Average prediction performance of the models, expressed by the RPIQ index,
ranged for our six models between 1.27 and 1.42.
Increasing the complexity of BRT models resulted in improving the prediction performance and 
the best R$^2$ value was obtained for the F$_g$ model with a value of 0.36.
 \\

Predictions with the LU$_g$ model exhibited, on average, limited bias (Fig.\ref{CVPlot}c). 
Important differences appeared when comparing mean prediction errors or mean root squared
errors to median prediction errors or median squared errors (Fig. \ref{CVPlot}c-f).	
This indicated a skewed distribution of errors. When assessed by the median of the 
error distribution (Fig. \ref{CVPlot}e), the geostatistical predictions are
shown to have a positive median-bias, whereas the standalone BRT predictions have
median-bias close to zero.
Similarly, the skewness of the distribution resulted in considerably larger root mean
squared errors (Fig. \ref{CVPlot}d),
with a lowest value of 2.83~kg/m$^2$ compared to root median squared errors
(Fig. \ref{CVPlot}f) with a lowest value of 1.43~kg/m$^2$.\\

\subsection{Performance comparisons of BRT models with or without spatial component}
For our French dataset, adding a spatial term to the models resulted in improvements in
terms of R2, RPIQ and the mean measures of prediction, RMSPE and MPE. These improvements
were not significant for the L/L$_g$ and F/F$_g$ model comparisons, for the  R2,
RPIQ and RMSPE.
However, in terms of the median measures, RMedSPE and MedPE, the standalone BRT predictions
generally gave the better results.


\begin{table*} 
\caption{Fitted variogram parameters in transformed units and cross-validation statistics.
 Mat\`{e}rn parameters: $C_0$ is the nugget variance, $C_1$ is the partial sill variance,  $\varphi$ is a spatial
 parameter expressed in km and $\kappa$ is a smoothness parameter.
$\overline{\theta}$ and $\breve{\theta}$ are the validation statistics before Winsorizing and 
$\overline{\theta}_w$ and $\breve{\theta}_w$ after Winsorizing (for the three models within the 95\% confidence interval). 
$N$ is the number of plots Winsorized and $c$ is the Winsorizing constant.}
\begin{tabular}{lcccccccccc}
  \hline
 & $C_{0}$ & $C_{1}$ & $\varphi$ & $\kappa$ & $\overline{\theta}$ & $\breve{\theta}$ & N & c & $\overline{\theta}_w$ & $\breve{\theta}_w$ \\ 
  \hline
LU$_g$ & 0.112 & 0.059 & 95.99 & 0.40 & 1.18 & 0.46 & 61 & 2.18 & 1.000 & 0.445 \\ 
  L$_g$ & 0.086 & 0.010 & 11.99 & 10.00 & 1.15 & 0.46 & 46 & 2.28 & 1.000 & 0.452 \\ 
  F$_g$ & 0.082 & 0.005 & 16.18 & 10.00 & 1.12 & 0.44 & 42 & 2.31 & 1.000 & 0.433 \\ 
   \hline
\end{tabular}
\label{varioFit}
\end{table*}


\begin{figure}
\includegraphics[width=95mm]{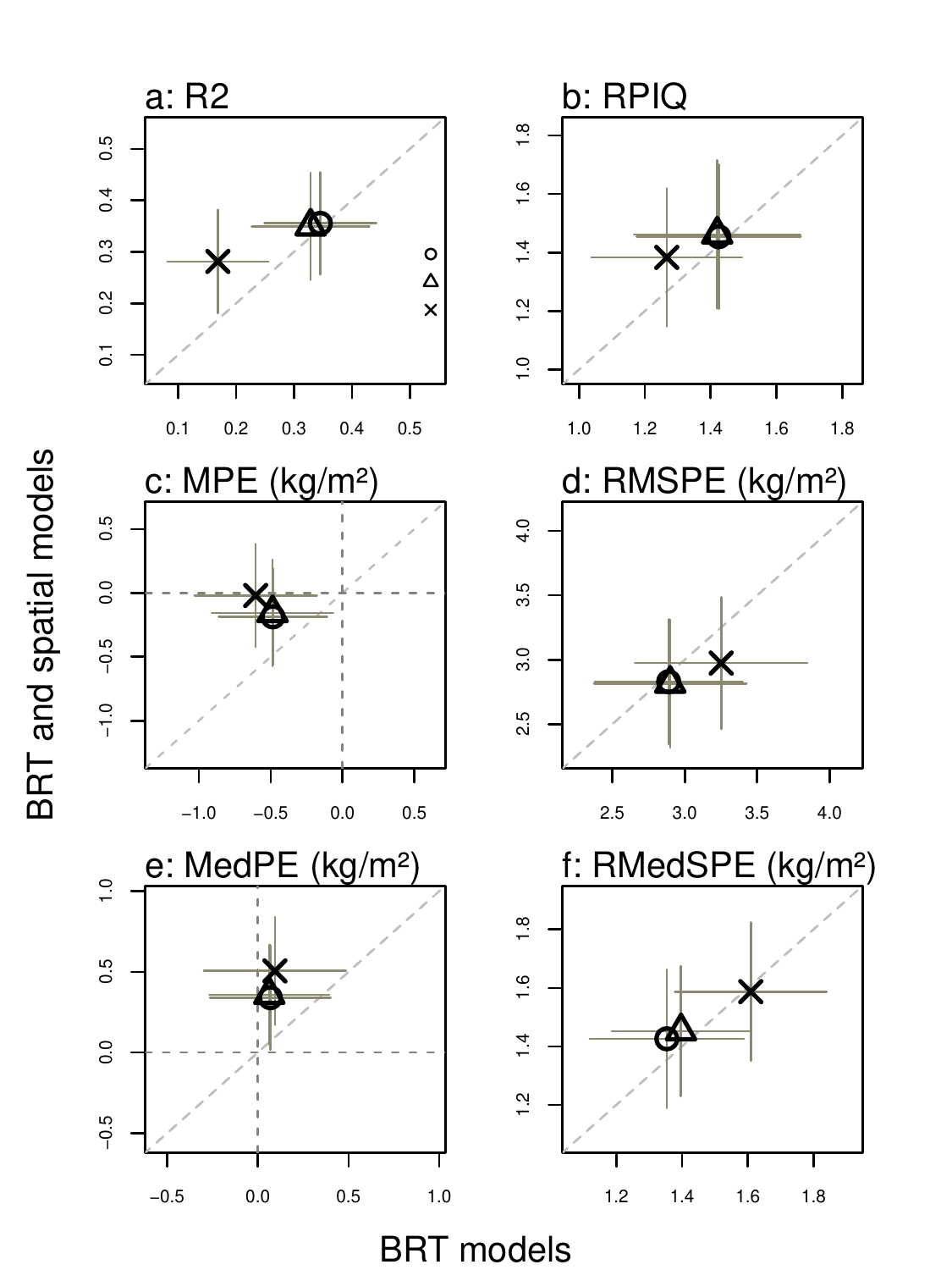}
\caption{
Performance of the six different models assessed using the 6 performance indices.
On each diagram, the values on the x-axis correspond to the aspatial models
(BRT only): the LU, L and F models. Values on the y-axis correspond to the LU, L and F 
models plus a spatial term, \emph{i.e.} the LU$_g$, L$_g$ and F$_g$ models. Horizontal and
vertical bars represent the 95\% confidence intervals around mean values
over the cross validation repetitions, for the BRT models only and the BRT 
with a spatial term models, respectively. 
The dotted lines correspond to the $y=x$ function and for the c and e diagrams the $y=0$ and 
the $x=0$ lines were added.}
\label{CVPlot}
\end{figure}


\begin{figure*}
\begin{center}
\includegraphics{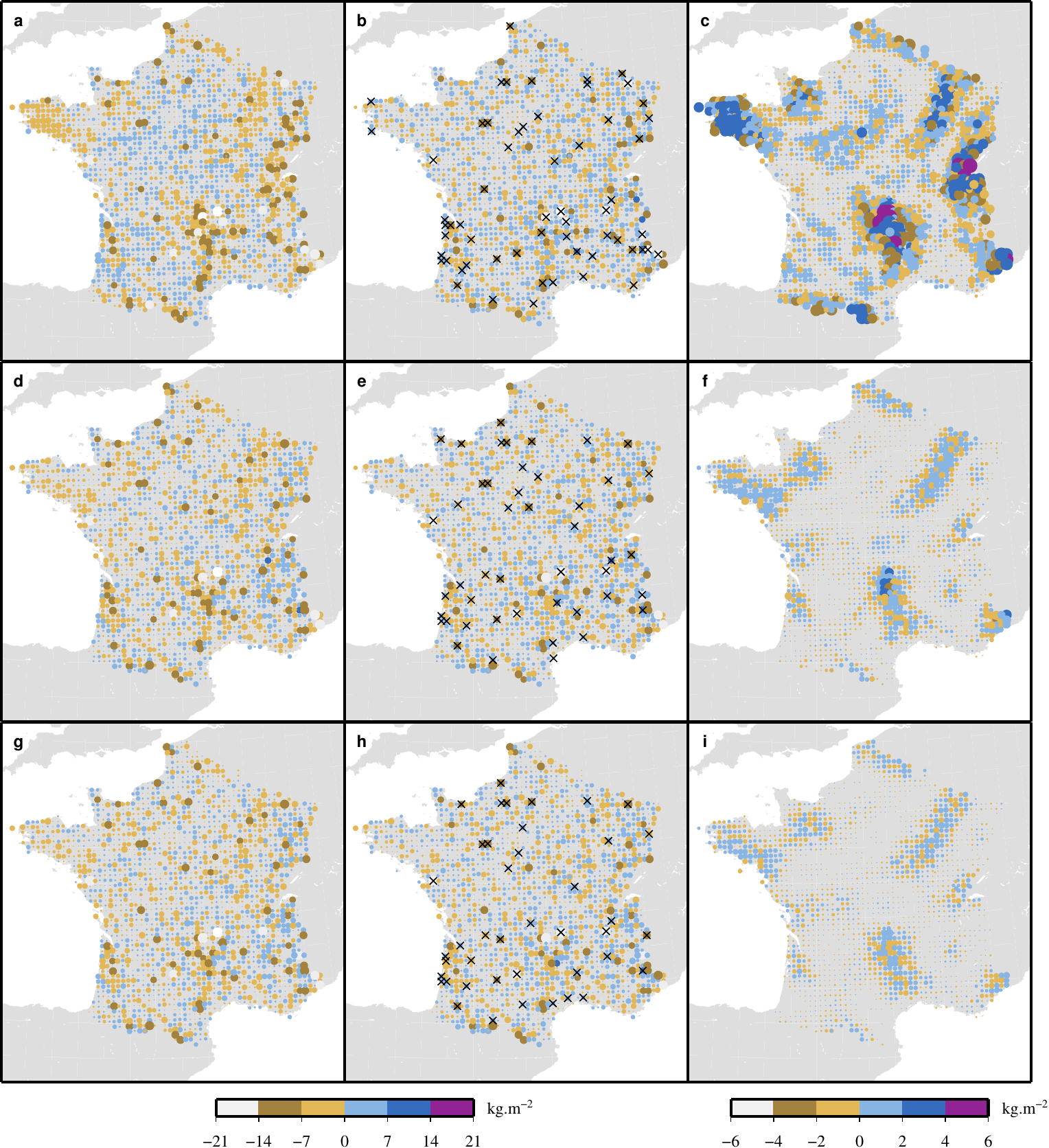}%
\caption{Average prediction error (predicted minus observed values) on each RMQS site over cross-validation repetitions where it was considered as independent data, with the LU, LU$_g$, L, L$_g$, F and F$_g$ models on maps a, b, d, e, g and h respectively. Positive values indicate a positive bias and \textit{vice versa}. Improvements from LU to LU$_g$, L to L$_g$ and F to F$_g$ models are given on maps c, f, and i respectively. For instance, map c gives the absolute error of the LU model minus the absolute error of the LU$_g$ model. Positive values indicate that adding a spatial component improved predictions at this location. Size of the dots is an increasing function of absolute errors (or absolute improvement for maps c, f and i). Crosses are outliers of spatial models fitted on the whole dataset.
}
\label{residMaps}
\end{center}
\end{figure*}

The improvement resulting from the addition of the spatial component was a
decreasing function of the complexity of the BRT model. 
This is shown clearly on Fig.\ref{CVPlot}a, b, d and f.
The R$^2$ for the LU model was improved from 0.17 to 0.28
when adding a spatial term. The map of errors (Fig.\ref{residMaps}a)
reveals regions where the LU model exhibited a strong negative bias,
such as south west Brittany (area 1; for reference of area numbers, see Fig. \ref{mape}),
and mountainous areas such as the Massif Central (area 5), Alps (area 4), and Vosges
on the eastern part of the French territory (area 3). In other regions, it exhibited a
positive bias, such as some of the parts of the south-west of France. 
The map of improvement between the LU and LU$_g$ models
(map of differences,  for each RMQS site, between the absolute errors of prediction with one BRT model and
the absolute error with its spatial counterpart, Fig.\ref{residMaps}c for the LU/LU$_g$ models)
shows areas with a dramatic improvement of predictions, and more specifically where the BRT predictions were
strongly biased.
It should be noted that the strongly biased predictions almost disappeared with the most complex model (model F, Fig.\ref{CVPlot}g).
Some under-estimations remained, although much smaller than  for the LU model, in western coastal areas.

Measured using the R$^2$ index, the improvement yielded by adding a spatial component to the F model
was not significant, with R$^2$ values going from 0.35 to 0.36.
Noticeably, the root of the median squared prediction errors exhibited a limited 
but significant degradation (from 1.35 to 1.43 for the F and the F$_g$ models, respectively).
The spatial distribution of improvements (Fig.\ref{residMaps}i) for this model
was clear for the south west Britanny region.
In many other areas the improvement was even more limited with some sites 
where prediction was improved and others where prediction was degraded (Fig.\ref{residMaps}i).
These were areas with high absolute errors (\emph{e.g.} the Massif Central Fig.\ref{residMaps}g).
Interestingly, there was no significant difference in the performance of the for F$_g$ and L$_g$
models. This result indicated that adding a spatial component to the intermediate BRT model yielded
similar results to adding a spatial component to the most complex model. 

\begin{table}[ht]
\centering
\begin{tabular}{rrrrrrr}
  \hline
 & LU & LUg & L & Lg & F & Fg \\ 
  \hline
RPIQ & 1.27 & 1.38 & 1.42 & 1.46 & 1.42 & 1.45 \\ 
  R2 & 0.17 & 0.28 & 0.33 & 0.35 & 0.34 & 0.35 \\ 
  RMSPE & 3.25 & 2.97 & 2.90 & 2.81 & 2.89 & 2.83 \\ 
  RMedSPE & 1.61 & 1.59 & 1.40 & 1.45 & 1.35 & 1.43 \\ 
  MPE & -0.60 & -0.02 & -0.49 & -0.16 & -0.48 & -0.19 \\ 
  MedPE & 0.09 & 0.51 & 0.06 & 0.36 & 0.07 & 0.34 \\ 
   \hline
\end{tabular}
\caption{Performance of the six different models assessed using the 6 performance indices.
Values given are the mean index values over the  200  repetitions of the
cross-validation procedure. All values but for the R$^2$ and RPIQ indices are given in kg/m$^2$.} 
\label{perfIndexes}
\end{table}

\section{Discussion}
%
%
\subsection{Spatial dependence of SOC stocks}


The spatial dependence of the BRT residuals decreased as the complexity of the BRT models was increased. 
The variogram parameters provide some information about SOC controlling factors not included in the BRT
 model. For instance, when land use and clay content is included (in the LU model), the correlation range of model
 residuals lies between 300 and 400 km (Fig.\ref{variograms}). This gives an indication of the
 correlation range of the most important SOC controlling factors missing from the LU model. 
Hence, when attempting to improve the LU model of SOC stocks spatial dependence, one should
 look for controlling factors whose correlation range is less than 300 to 400 km.
The L model included other controlling factors, such as clay content, which decreased
both the total variance of residuals and their correlation range,
to around 100 to 200~km. Lastly, the F model handled 
most of the spatial dependence by including three more drivers, the pH, the parent material and the regional 
statistics regarding organic matter additions. 
However, the high nugget in the variograms from the residuals of each BRT model,
including the F model, indicated
that other controlling factors greatly influence SOC spatial distributions at ranges below the resolution
of our dataset, \textit{i.e.} 16~km. This is consistent with the results of many other studies.
For instance, in \citet{Ungar2010}, the residuals of a model of SOC (\%) taking into account administrative zonation
and soil functional types were analysed by variography. They also found that most of the
spatially structured variance
was accounted for by a short range component (in their case 1500-2000~m). 
Another possible explanation for the high
nugget could be that the uncertainty attached to most of the covariates (drivers)
maps is high, especially for the covariates derived from the 1:1000~000 soil map.
%
\subsection{Assessing the performance of one single model}
It is difficult to draw conclusions regarding the performance of the present models compared to those of other studies dealing with SOC prediction and mapping. 
Some deal with SOC contents when other deal with SOC densities or stocks. When
working on SOC stocks, the bulk densities are required, and if these are estimated
(rather than measured), then the methodology
for estimating bulk density might have great consequences  \citep{Liebens2003}. Many studies use pedotransfer
functions (PTFs) for estimating bulk density without accounting for the associated errors \citep{Schrumpf2011}.
\citet{Ungar2010} estimated through a Monte Carlo approach
that uncertainty resulting from their PTF ranged between 0.55 and 7.72~T ha$^{-2}$ 
depending on the SOC content. \cite{Schrumpf2011} showed that the use of PTFs for
estimating bulk densities can lead to wrong or biased estimates of SOC stocks. 
However it is currently not entirely clear to what extent measuring bulk densities is worth, considering the cost. 
This cost could alternatively be used to collect further SOC concentration data and thus improve calibration 
and validation datasets. 
Comparison between the studies is also made more complex by the differences between
validation procedures (validation with an independent dataset, k-fold cross-validation, leave-one-out
cross-validation). Furthermore, as quoted by \citet{Minasny2013} and
\citet{Grunwald2009}, it is quite common that validation of SOC model predictions is missing entirely from a
study. 
The best models presented in our study (the F and F$_g$ models, see Fig.\ref{CVPlot})
performed comparably to
those of \citet{Lo_Seen2010}, fitted on soil data from the Western Ghats biodiversity hotspot (India). The models yielded,
using a cross-validation scheme similar to the one applied here, RMSPE of 2.6~kg~m$^{-2}$ and R$^2$ of 0.45, to be 
compared to the RMSPE of 2.83~kg~m$^{-2}$ and R$^2$ of 0.36 obtained here for the F$_g$ model, along
with a MPE value of -0.19 kg~m$^{-2}$. Considering
national SOC prediction, \citet{Phachomphon2010} produced 0-100~cm estimates using
inverse distance weighting with 12 neighbours and ordinary cokriging, 
yielding MPEs of -0.2 and -0.1~kg~m$^{-2}$   and a RMSPE of 2.2 and 2.1~kg~m$^{-2}$,
respectively.
\citet{Mishra2009} produced estimates, for the Indiana state (USA) with a MPE  of -0.59~kg~m$^{-2}$
and RMSPE of 2.89~kg~m$^{-2}$. This study involved the fitting of SOC depth distributions, as 
recently proposed by \citet{Kempen2011}. This last study, is among the most successful, with a 
rigorous validation scheme and an moderate extent (125~km$^2$), giving an R$^2$ of 0.75 for prediction
on independent locations. Other studies on areas of the same order of
magnitude as the area of the French territory ($i.e.$ >50~000~km$^2$) are referenced in the comprehensive
review of \citet{Minasny2013}. The R$^2$ values of our models
are towards the lower end of R$^2$ values in studies mentioned in this review. Their
performance is also remarkably lower compared to similar models
presented in \citet{bg-8-1053-2011} and \citet{Meersmans2012}. 
This drop in model performance is likely a result of the
uncertainty of the clay content estimated from the 1:1000~000 soil map (as compared to 
the measured clay contents used in the previous studies). This is indicated by the
importance (quantified using the BRT variable importance index) of clay related
variables in the L and F models of the present study. These variables ranked  at best
 7$^{th}$ and 7$^{th}$
in the L and F models, respectively. This is to be compared to the first rank
obtained by the $clay$ variable in the $Extra$ model presented in
\citet{bg-8-1053-2011}, fitted and validated with measured clay contents.
Thus, for the two previous studies for the French territory,
the model performance for mapping might have been
overestimated because some variables used for validation were observed at site level.
In the present study, models are validated using data
estimated from ancillary maps, providing a more realistic assessment of model
performance for mapping. Such small differences in the model validation schemes
are difficult to trace and might further complicate comparison between different
studies.

\subsection{Distribution of predictions errors}
Another issue worth commenting on here is the distribution of SOC stocks
predictions errors. 
BRT modelling of log transformed SOC stocks gave residuals that were close to normal, with outliers.
These residuals were modelled using a robust geostatistical approach, and a back transformation 
proposed for log-normal ordinary kriging was applied.
The final predictions exhibited a limited bias (MPE=-0.19~kg~m$^{-2}$ for the F$_g$
model, Table \ref{perfIndexes}), a problem that can arise in lognormal kriging due to the sensitivity of the back-transform
 to the variogram parameters and to the assumption of a lognormal distribution \citep{Webster2007}. 
Although we currently have no ready solution for providing unbiased predictions,
especially for the L$_g$ and F$_g$ models, we note that the MPE is small in comparison 
to the RMSPE (less than 5 \% of the RMSPE), which compares favourably with results of other studies reported above.
Without the spatial component, the BRT predictions (back-transformed with a simple exponential,
see Algorithm 1), showed negative mean-bias (i.e. under-prediction on average).
This is logical because the BRT method ensures unbiased
predictions on its predicted variable, here the log transformed SOC stocks; therefore, 
back-transformation of the BRT predictions through the exponential function results 
in a negative mean-bias for SOC stocks on the original scale.

Further insight is provided by examining the behaviour of other performance
indices, such as the median prediction error or the median squared prediction
error.
The lognormal kriging back-transformation aims to provide mean-unbiased predictions on the original
 scale, hence the reasonably small MPE. However, with a skewed distribution of errors, the predictions
 cannot also be made to be median-unbiased, hence the MedPE of the geostatistical predictions
 is positive. Without the geostatistical component, the back-transform of the BRT predictions
 (through the exponential function) preserves the median-unbiased property, giving low values
 of MedPE, but introduces mean-bias. Comparisons between the results of our BRT predictions
 and their spatial counterparts should be made with this in mind; the differences could be at
 least partly due to the different objectives of the back-transformed predictors. 
Since SOC distributions are most commonly 
as log-normal, prediction error distributions are also skewed, and perhaps these
further measures (MedPE and RMedPE, which are robust
to extreme prediction errors at a small number of locations) can add useful information about model performance.

We note here that the BRT approach could be applied to model the SOC stocks directly, without
 the need for any transformation (as shown by \citealp{bg-8-1053-2011}). We would expect the resulting
 predictions to have low MPE, but a positive MedPE (a similar pattern to the results of the
 geostatistical approach). We tested this direct BRT modelling approach with the F BRT model;
 mean prediction errors were improved from -0.48 to 0.01~kg~m$^{-2}$,
 whilst median prediction errors were
 increased from 0.07 to 0.45~kg~m$^{-2}$.
 In terms of squared errors, the RMSPE improved slightly from
 2.89 to 2.82~kg~m$^{-2}$, 
whilst the RMedSPE increased from 1.35 to 1.5~kg~m$^{-2}$.
In this work, we applied BRT
 modelling to the log-transformed data so that residuals would be approximately normal, thus
 allowing the robust geostatistical approach to be applied. However, if all that was required
 was predictions of the SOC stock through a BRT approach, then it may be better to model the 
raw SOC stock data directly.


\subsection{Relevance of the models for SOC mapping}

Models comparisons enable one to come up with recommendations regarding the best models 
for assessing a specific question. Of course, the quality of the models should be assessed using several criteria
as the question of interest is asked within a specific context (data availability, nature of the considered systems, available
statistical and modelling knowledge, computing cost). Several comparison criteria may be defined : the 
Several comparison criteria may be defined : the 
technical knowledge (Know-Q) and the pedological knowledge (Know-P)  needed for fitting,
validating and applying models \citep{Grunwald2009}. 
We may add a criterion related to the nature of the required datasets, again, for fitting, 
validating and applying models, and another one related to the performance of the models, assessed through validation
procedures. Although other criteria might be defined,  those might be considered as the main ones for predictive models.
The best models would be those which, given the available Know-Q, Know-P and the datasets, yield the best performance.

Several studies of SOC mapping include model comparison in order to provide the best performing model and advices regarding which
model should be used in a specific context. Comparing the results of these different studies is not straightforward since the 
pedological contexts change from one study to another.
In studies based on the application of geostatistics, model comparison
is usually carried out by comparing simple geostatistical models with more advanced
approaches designed to incorporate covariate data (e.g. cokriging,
\citealp{MCBRATNEY1983}, linear mixed models \citealp{Lark2004}, or more generally
scorpan-kriging models \citealp{McBratney2003}).
The conclusion is consistently that including variables representing SOC drivers in geostatistical models improves model
performance \citep[for instance see][]{Kempen2011, Vasques2010, Ungar2010}. The cost of such an improvement is that it
leads to an increase of the Know-P and the Know-Q. On one hand, such models might involve a great 
amount of technicality. On the other hand, the availability at observational sites of information regarding 
the included drivers
is then also required  for the fit, the validation and later on the prediction. \\

Fewer studies considered the question the other way around by including geostatistics in regression-based scorpan
 models, such as the BRT models considered here or by comparing regression models to regression-kriging models. 
On a 187,693 km$^2$ area, \citet{Zhao2010} showed that simple regression trees (RT) exhibited the best performance
 when compared to regression kriging and artificial neural network-kriging, among other methods. They concluded
 that their predictive models mostly rely on their ability to integrate secondary information into spatial 
prediction. In our case, the conclusions are contrasted. The LU$_g$ model applied a robust geostatistical approach
 to the residuals of the simplest BRT model (the LU model, which included land use and clay content as the only
 fixed effects, among the most important SOC drivers at the national scale of France,
\citealp{bg-8-1053-2011}).
 This approach exhibited comparable but lower performance, in terms of R$^2$, RPIQ, and RMSPE compared to the 
more complex regression models (L and F) processing all the available ancillary data. Therefore, we conclude
 that adding a spatial component to a simple regression model can give similar improvements to adding more 
predictors to the model.

Unbiased predictions might be achieved either by BRT
modelling on the original scale \citep[as shown by][]{bg-8-1053-2011} or by BRT
modelling of the log-transformed response and applying a geostatistical
treatment. When it comes to mapping, one may wonder if preserving the mean of
SOC stock distributions is more important than preserving the median. The mean might
be more imporant in order to report total SOC stocks at the national scale, but preserving the
median might result in more realistic maps. It is essentially a modelling choice, as to whether
mean-unbiasedness or median-unbiasedness is required. 

\subsection{Further recommendations for SOC mapping at the national scale}
Our best model (the F$_g$ model) only explained 36\% of the SOC variation. 
It is possible that local kriging methods, rather than the global kriging applied here, 
could lead to improved predictions in some areas, although the choice of appropriate local 
neighbourhood sizes then provides an additional issue.  Other regression models could be tested,
such as support vector machines (SVM), random forests \citep{hastie2001} or the
Cubist modelling approach (\textit{e.g.} \citealp{Bui2009}). 
These models could result in different residual distributions but in our opinion,
the consequences on the performance of their spatial couterpart are likely to
be limited. Some of them, such as SVM require more technical knowledge, thus increasing the Know-Q factor, 
compared to the BRT models proposed here, for which efficient working guides have been proposed \citep{Elith2008}.
\citet{Grunwald2009} stated that the future improvements in the prediction of soil properties
does not rely on more 
sophisticated quantitative methods, but rather on gathering more useful and higher quality data.
Choosing between gathering more data or improving the modelling is indeed the choice modellers are
facing when attempting to improve SOC maps. We show here that the choice might not be
as straightforward as stated by \citet{Grunwald2009} : at the national scale, even a
simple model based solely on landuse and clay, when complemented by geostatistics,
performed comparably to a model where all the available ancillary data was included
(for France, at the time of the study, these were land use, soil, climate and npp
maps). Therefore, for a country where only landuse and clay maps were available, the most
efficient way to improve predictions in the short term would certainly be to consider
geostatistical modelling of residuals (\textit{i.e.} improving the modelling, rather
than gathering new ancillary data). 
Furthermore, other datasets, on the same extent (\textit{i.e.} national extent)
but with different resolution might be more suited to geostatistics. Here,
the 16x16km$^2$ does not allow for modelling spatial autocorrelations
occurring at small scales. Many studies have demonstrated such an autocorrelation when more "local"
neighbourhoods can be studied (\citealp{Mabit2010, Don2007, Rossi2009, Wang2009, Spielvogel2009} 
with an extent <50km$^2$ and  \citealp{Mishra2009} at coarser extents and using a
non-systematic sampling scheme).\\
On the other hand, adding spatial terms to the most complex models only increased know-Q
to our data-analysis scheme. 
More generally, the higher the uncertainty in maps of ancillary variables, the more likely it is that models based solely on 
SOC spatial dependency or including only few good quality (in terms of data uncertainty) predictors will outperform
complex models using many ancillary variables. 


For France, other SOC predictors could be included in our regression models,
and result in significant improvements. There are different possibilities \citep{bg-8-1053-2011}, but of course, these
improvements depend on the increase in Know-P and data-collection one is willing to consider.
Having a better soil map is obviously a very good candidate. This is exemplified here
by the drop in the F model performance between the present study and the work by
\citet{bg-8-1053-2011}. \\

It is also worth noting that an advantage of using multiple regression tools, such as
the BRT models, comes from studying the fitted relationships between the response 
and the predictors, which may in turn bring 
additional knowledge. For instance, BRT was used in \citet{bg-8-1053-2011} to rank the effects
of the SOC stocks driving factors.



\section{Conclusion}
Based on the results of the present study, and others found in the literature, we formulate the following recommendations.
 These recommendations apply for France but the French diversity in terms of pedoclimatic conditions might make these
 recommendations valid for other countries as well. If the information contained in the relationships between
 the ancillary variables and the SOC stocks are strong enough, then standalone robust regression models such as
 BRT - which enable one to take into account in a flexible way non-linearities and interactions exhibited by the
 datasets - could prove sufficient for SOC mapping at the national scale. This conclusion is valid provided that i)
 care is exercised in model fitting \citep{Elith2008} and validating, ii) the dataset does not allow for modelling
 local spatial autocorrelations, as it is the case for many national systematic sampling schemes, and iii) the ancillary
 data are of suitable quality. However, the results in this paper demonstrate that it should also be prudent to use
 geostatistical  methods to check for spatial autocorrelation in the BRT residuals. If found, which was the case
 for the simpler  of our BRT models (which failed to capture all the important SOC drivers at a
 national scale), then a kriging approach
 applied to the BRT residuals can provide a more accurate map of SOC stocks. Furthermore, even if the spatial correlation
 fails to significantly improve SOC predictions globally, it is possible that by mapping the BRT model residuals we
 can highlight regional errors in the BRT model, and thereby provide information to guide research into further SOC
 model development.

\section*{Acknowledgements}
The sampling and soil analyses were supported by a French Scientific Group of Interest 
on soils: the GIS Sol, involving the French Ministry of Ecology, Sustainable Development
and Energy (MEDDE), the French Ministry of Agriculture, Food and Forestry (MAAF), the
French Agency for Environment and Energy Management (ADEME), the Institute for Research
and Development (IRD), the National Institute
of Geographic and Forest Information (IGN) and the National Institute
for Agronomic Research (INRA). This work was supported by the EU project ``Greenhouse
gas management in European land use systems (GHG-Europe)''
(FP7-ENV-2009-1-244122. The authors thank all the soil surveyors and technical
assistants involved in sampling the sites. Special thanks are addressed to the technical
assistants from the National French Soil Bank for sample handling and preparation. 
\section*{References}

\bibliographystyle{elsarticle-harv}
\bibliography{bilbio}

\begin{thebibliography}{71}
\expandafter\ifx\csname natexlab\endcsname\relax\def\natexlab#1{#1}\fi
\expandafter\ifx\csname url\endcsname\relax
  \def\url#1{\texttt{#1}}\fi
\expandafter\ifx\csname urlprefix\endcsname\relax\def\urlprefix{URL }\fi

\bibitem[{ADEME(2007)}]{ADEME2007}
ADEME, 2007. Bilan des flux de contaminants entrant sur les sols agricoles de
  france métropolitaine: bilan qualitatif de la contamination par les éléments
  tracés métalliques et les composés tracés organiques et application
  quantitative pour les éléments tracés métalliques. Tech. rep., (French
  Environment and Energy Management Agency).

\bibitem[{Arrouays et~al.(2001)Arrouays, Deslais, and Badeau}]{48}
Arrouays, D., Deslais, W., Badeau, V., 2001. The carbon content of topsoil and
  its geographical distribution in france. Soil Use and Management 17~(1),
  7--11.

\bibitem[{Arrouays et~al.(2002)Arrouays, Jolivet, Boulonne, Bodineau, Saby, and
  Grolleau}]{Arrouays2002}
Arrouays, D., Jolivet, C., Boulonne, L., Bodineau, G., Saby, N., Grolleau, E.,
  2002. A new initiative in france: a multi-institutional soil quality
  monitoring network. Comptes rendus de l'Academie d'Agriculture de France
  88~(5).

\bibitem[{Batjes(1996)}]{Batjes1996}
Batjes, N.~H., 1996. Total carbon and nitrogen in the soils of the world.
  European Journal of Soil Science 47~(2), 151--163.

\bibitem[{Bellon-Maurel et~al.(2010)Bellon-Maurel, Fernandez-Ahumada, Palagos,
  Roger, and McBratney}]{Bellon-Maurel2010}
Bellon-Maurel, V., Fernandez-Ahumada, E., Palagos, B., Roger, J.-M., McBratney,
  A., 2010. Critical review of chemometric indicators commonly used for
  assessing the quality of the prediction of soil attributes by nir
  spectroscopy. Trac-Trends in Analytical Chemistry 29~(9), 1073--1081.

\bibitem[{Bivand et~al.(2008)Bivand, Pebesma, and Gómez-Rubio}]{Bivand2008}
Bivand, R.~S., Pebesma, E.~J., Gómez-Rubio, V., 2008. Applied Spatial Data
  Analysis with R. UseR! Springer.

\bibitem[{Breiman et~al.(1984)Breiman, Friedman, Olshen, and Stone}]{Brei1984}
Breiman, L., Friedman, J.~H., Olshen, R.~A., Stone, C.~J., 1984. Classification
  and regression trees. Wadsworth, Inc. Monterey, Calif., U.S.A.

\bibitem[{Bui et~al.(2009)Bui, Henderson, and Viergever}]{Bui2009}
Bui, E., Henderson, B., Viergever, K., 2009. Using knowledge discovery with
  data mining from the australian soil resource information system database to
  inform soil carbon mapping in australia. Global Biogeochemical Cycles 23.

\bibitem[{Chaplot et~al.(2009)Chaplot, Bouahom, and Valentin}]{CHAPLOT2009}
Chaplot, V., Bouahom, B., Valentin, C., 2009. Soil organic carbon stocks in
  laos: spatial variations and controlling factors. Global Change Biology
  16~(4), 1380--1393.

\bibitem[{Chunfaand et~al.(2009)Chunfaand, Jiapingand, Yongmingand, and
  Liminand}]{Wu2009}
Chunfaand, W., Jiapingand, L., Yongmingand, Z., Liminand, DeGloria, S.~D.,
  2009. Spatial prediction of soil organic matter content using cokriging with
  remotely sensed data. Soil Science Society of America Journal 73~(4),
  1202--1208.

\bibitem[{Coleman et~al.(1997)Coleman, Jenkinson, Crocker, Grace, Klir,
  Korschens, Poulton, and Richter}]{coleman1997}
Coleman, K., Jenkinson, D.~S., Crocker, G.~J., Grace, P.~R., Klir, J.,
  Korschens, M., Poulton, P.~R., Richter, D.~D., 1997. Simulating trends in
  soil organic carbon in long-term experiments using rothc-26.3. Geoderma
  81~(1-2), 29--44.

\bibitem[{Don et~al.(2007)Don, Schumacher, Scherer-Lorenzen, Scholten, and
  Schulze}]{Don2007}
Don, A., Schumacher, J., Scherer-Lorenzen, M., Scholten, T., Schulze, E.-D.,
  2007. Spatial and vertical variation of soil carbon at two grassland sites -
  implications for measuring soil carbon stocks. Geoderma 141~(3-4), 272--282.

\bibitem[{Dowd(1984)}]{Dowd1984}
Dowd, P.~A., 1984. The variogram and kriging: robust and resistant estimators.
  In: Verly, G., David, M., Journel, A.~G., Marechal, A. (Eds.), Geostatistics
  for Natural Resources Characterrization. D. Reidel, Dordrecht, pp. 91--106.

\bibitem[{Elith et~al.(2008)Elith, Leathwick, and Hastie}]{Elith2008}
Elith, J., Leathwick, J.~R., Hastie, T., 2008. A working guide to boosted
  regression trees. Journal of Animal Ecology 77~(4), 802--813.

\bibitem[{Eswaran et~al.(1993)Eswaran, Vandenberg, and Reich}]{Eswaran1993}
Eswaran, H., Vandenberg, E., Reich, P., 1993. Organic-carbon in soils of the
  world. Soil Science Society of America Journal 57~(1), 192--194.

\bibitem[{Freund and Schapire(1996)}]{Freund96experimentswith}
Freund, Y., Schapire, R.~E., 1996. Experiments with a new boosting algorithm.
  In: Machine Learning: Proceedings of the Thirteenth International Conference.
  Morgan Kauffman, San Francisco., pp. 148--156.

\bibitem[{Friedman(2001)}]{Friedman2001}
Friedman, J.~H., 2001. Greedy function approximation: A gradient boosting
  machine. Annals of Statistics 29~(5), 1189--1232.

\bibitem[{Goovaerts and Kerry(2010)}]{Goov2010}
Goovaerts, P., Kerry, R., 2010. Using ancillary data to improve prediction of
  soil and crop attributes in precision agriculture. In: Oliver, M.~A. (Ed.),
  Geostatistical Applications for Precision Agriculture. Springer Science, pp.
  167--194.

\bibitem[{Grimm et~al.(2008)Grimm, Behrens, Marker, and Elsenbeer}]{Grimm2008}
Grimm, R., Behrens, T., Marker, M., Elsenbeer, H., 2008. Soil organic carbon
  concentrations and stocks on barro colorado island - digital soil mapping
  using random forests analysis. Geoderma 146~(1-2), 102--113.

\bibitem[{Grunwald(2009)}]{Grunwald2009}
Grunwald, S., 2009. Multi-criteria characterization of recent digital soil
  mapping and modeling approaches. Geoderma 152~(3-4), 195--207.

\bibitem[{Hastie et~al.(2001)Hastie, Tibshirani, and Friedman}]{hastie2001}
Hastie, T., Tibshirani, R., Friedman, J., 2001. The Elements of Statistical
  Learning, Data Mining, Inference, and Prediction, Second Edition. Springer
  Series in Statistics.

\bibitem[{Hawkins and Cressie(1984)}]{Hawkins1984}
Hawkins, D.~M., Cressie, N., 1984. Robust kriging - a proposal. Journal of the
  International Association for Mathematical Geology 16~(1), 3--18.

\bibitem[{Kempen et~al.(2011)Kempen, Brus, and Stoorvogel}]{Kempen2011}
Kempen, B., Brus, D.~J., Stoorvogel, J.~J., 2011. Three-dimensional mapping of
  soil organic matter content using soil type-specific depth functions.
  Geoderma 162~(1-2), 107--123.

\bibitem[{Kerry et~al.(2012)Kerry, Goovaerts, Rawlins, and
  Marchant}]{Kerry2012}
Kerry, R., Goovaerts, P., Rawlins, B.~G., Marchant, B.~P., 2012. Disaggregation
  of legacy soil data using area to point kriging for mapping soil organic
  carbon at the regional scale. Geoderma 170, 347--358.

\bibitem[{King et~al.(1995)King, Daroussin, Le~Bas, Tavernier, and van
  Ranst}]{King1995}
King, D., Daroussin, J., Le~Bas, C., Tavernier, R., van Ranst, E., 1995. The eu
  soil geographical database. EUR 16232 EN. Office for Official Publications of
  the European Communities Luxembourg.

\bibitem[{Lacarce et~al.(2012)Lacarce, Saby, Martin, Marchant, Boulonne,
  Meersmans, Jolivet, Bispo, and Arrouays}]{Lacarce2012}
Lacarce, E., Saby, N. P.~A., Martin, M.~P., Marchant, B.~P., Boulonne, L.,
  Meersmans, J., Jolivet, C., Bispo, A., Arrouays, D., 2012. Mapping soil pb
  stocks and availability in mainland france combining regression trees with
  robust geostatistics. Geoderma 170, 359--368.

\bibitem[{Lal(2004)}]{Lal2004}
Lal, R., 2004. Soil carbon sequestration to mitigate climate change. Geoderma
  123~(1-2), 1--22.

\bibitem[{Lark and Cullis(2004)}]{Lark2004}
Lark, R.~M., Cullis, B.~R., 2004. Model-based analysis using reml for inference
  from systematically sampled data on soil. Eur J Soil Sci 55~(4), 799--813.

\bibitem[{Lawrence et~al.(2004)Lawrence, Bunn, Powell, and
  Zambon}]{Lawrence2004}
Lawrence, R., Bunn, A., Powell, S., Zambon, M., 2004. Classification of
  remotely sensed imagery using stochastic gradient boosting as a refinement of
  classification tree analysis. Remote Sensing of Environment 90~(3), 331--336.

\bibitem[{Lemercier et~al.(2006)Lemercier, Arrouays, Follain, Saby, Schvartz,
  and Walter}]{Lemercier2006}
Lemercier, B., Arrouays, D., Follain, S., Saby, N. P.~A., Schvartz, C., Walter,
  C., 2006. Broad-Scale Soil Monitoring Through a Nationwide Soil-Testing
  Database. Springer, Netherlands, pp. 273--281.

\bibitem[{Liebens and VanMolle(2003)}]{Liebens2003}
Liebens, J., VanMolle, M., 2003. Influence of estimation procedure on soil
  organic carbon stock assessment in flanders, belgium. Soil Use And Management
  19~(4), 364--371.

\bibitem[{Lo~Seen et~al.(2010)Lo~Seen, Ramesh, Nair, Martin, Arrouays, and
  Bourgeon}]{Lo_Seen2010}
Lo~Seen, D., Ramesh, B.~R., Nair, K.~M., Martin, M., Arrouays, D., Bourgeon,
  G., 2010. Soil carbon stocks, deforestation and land-cover changes in the
  western ghats biodiversity hotspot (india). Glob Change Biol 16~(6),
  1777--1792.

\bibitem[{Mabit and Bernard(2010)}]{Mabit2010}
Mabit, L., Bernard, C., 2010. Spatial distribution and content of soil organic
  matter in an agricultural field in eastern canada, as estimated from
  geostatistical tools. Earth Surface Processes And Landforms 35~(3), 278--283.

\bibitem[{Marchant et~al.(2010)Marchant, Saby, Lark, Bellamy, Jolivet, and
  Arrouays}]{Marchant2010}
Marchant, B.~P., Saby, N. P.~A., Lark, R.~M., Bellamy, P.~H., Jolivet, C.~C.,
  Arrouays, D., 2010. Robust analysis of soil properties at the national scale:
  cadmium content of french soils. European Journal of Soil Science 61~(1),
  144--152.

\bibitem[{Martin et~al.(2009)Martin, Lo~Seen, Boulonne, Jolivet, Nair,
  Bourgeon, and Arrouays}]{Martin2009}
Martin, M.~P., Lo~Seen, D., Boulonne, L., Jolivet, C., Nair, K.~M., Bourgeon,
  G., Arrouays, D., 2009. Optimizing pedotransfer functions for estimating soil
  bulk density using boosted regression trees. Soil Science Society of America
  Journal 73~(2), 485--493.

\bibitem[{Martin et~al.(2011)Martin, Wattenbach, Smith, Meersmans, Jolivet,
  Boulonne, and Arrouays}]{bg-8-1053-2011}
Martin, M.~P., Wattenbach, M., Smith, P., Meersmans, J., Jolivet, C., Boulonne,
  L., Arrouays, D., 2011. Spatial distribution of soil organic carbon stocks in
  france. Biogeosciences 8~(5), 1053--1065.
\newline\urlprefix\url{http://www.biogeosciences.net/8/1053/2011/}

\bibitem[{McBratney et~al.(2003)McBratney, Santos, and Minasny}]{McBratney2003}
McBratney, A.~B., Santos, M. L.~M., Minasny, B., 2003. On digital soil mapping.
  Geoderma 117~(1-2), 3--52.

\bibitem[{McBratney and Webster(1983)}]{MCBRATNEY1983}
McBratney, A.~B., Webster, R., 1983. Optimal interpolation and isarithmic
  mapping of soil properties .5. co-regionalization and multiple sampling
  strategy. Journal of soil Science 34~(1), 137--162.

\bibitem[{Meersmans et~al.(2008)Meersmans, De~Ridder, Canters, De~Baets, and
  Van~Molle}]{Meersmans2008}
Meersmans, J., De~Ridder, F., Canters, F., De~Baets, S., Van~Molle, M., 2008. A
  multiple regression approach to assess the spatial distribution of soil
  organic carbon (soc) at the regional scale (flanders, belgium). Geoderma
  143~(1-2), 1--13.

\bibitem[{Meersmans et~al.(2012)Meersmans, Martin, Lacarce, De~Baets, Jolivet,
  Boulonne, Lehmann, Saby, Bispo, and Arrouays}]{Meersmans2012}
Meersmans, J., Martin, M., Lacarce, E., De~Baets, S., Jolivet, C., Boulonne,
  L., Lehmann, S., Saby, N., Bispo, A., Arrouays, D., 2012. A high resolution
  map of french soil organic carbon. Agronomy for Sustainable Development,
  1--1110.1007/s13593-012-0086-9.
\newline\urlprefix\url{http://dx.doi.org/10.1007/s13593-012-0086-9}

\bibitem[{Minasny et~al.(2013)Minasny, McBratney, Malone, and
  Wheeler}]{Minasny2013}
Minasny, B., McBratney, A.~B., Malone, B.~P., Wheeler, I., 2013. Digital
  mapping of soil carbon. advances in agronomy, Vol 118 118, 1--47.

\bibitem[{Mishra et~al.(2009)Mishra, Lal, Slater, Calhoun, Liu, and
  Van~Meirvenne}]{Mishra2009}
Mishra, U., Lal, R., Slater, B., Calhoun, F., Liu, D.~S., Van~Meirvenne, M.,
  2009. Predicting soil organic carbon stock using profile depth distribution
  functions and ordinary kriging. Soil Science Society of America Journal
  73~(2), 614--621.

\bibitem[{Phachomphon et~al.(2010)Phachomphon, Dlamini, and
  Chaplot}]{Phachomphon2010}
Phachomphon, K., Dlamini, P., Chaplot, V., 2010. Estimating carbon stocks at a
  regional level using soil information and easily accessible auxiliary
  variables. Geoderma 155~(3-4), 372--380.

\bibitem[{Post et~al.(1982)Post, Emanuel, Zinke, and Stangenberger}]{POST1982}
Post, W.~M., Emanuel, W.~R., Zinke, P.~J., Stangenberger, A.~G., 1982. Soil
  carbon pools and world life zones. Nature 298~(5870), 156--159.

\bibitem[{Quintana-Segui et~al.(2008)Quintana-Segui, Le~Moigne, Durand, Martin,
  Habets, Baillon, Canellas, Franchisteguy, and Morel}]{Quintana-Segui2008}
Quintana-Segui, P., Le~Moigne, P., Durand, Y., Martin, E., Habets, F., Baillon,
  M., Canellas, C., Franchisteguy, L., Morel, S., 2008. Analysis of
  near-surface atmospheric variables: Validation of the safran analysis over
  france. Journal of Applied Meteorology And Climatology 47~(1), 92--107.

\bibitem[{{R Core Team}(2013)}]{RCore}
{R Core Team}, 2013. R: A Language and Environment for Statistical Computing. R
  Foundation for Statistical Computing, Vienna, Austria.
\newline\urlprefix\url{http://www.R-project.org/}

\bibitem[{Rawlins et~al.(2009)Rawlins, Marchant, Smyth, Scheib, Lark, and
  Jordan}]{Rawlins2009}
Rawlins, B.~G., Marchant, B.~P., Smyth, D., Scheib, C., Lark, R.~M., Jordan,
  C., 2009. Airborne radiometric survey data and a dtm as covariates for
  regional scale mapping of soil organic carbon across northern ireland.
  European Journal of Soil Science 60~(1), 44--54.

\bibitem[{Ribeiro and Diggle(2001)}]{Ribeiro2001}
Ribeiro, P.~J., Diggle, P.~J., 2001. geor: a package for geostatistical
  analysis. R-News 1~(2), 15--18.

\bibitem[{Ridgeway(2006)}]{Ridgeway2006}
Ridgeway, G., 2006. gbm: Generalized boosted regression models. r package
  version 2.1.

\bibitem[{Rossi et~al.(2009)Rossi, Govaerts, De~Vos, Verbist, Vervoort, Poesen,
  Muys, and Deckers}]{Rossi2009}
Rossi, J., Govaerts, A., De~Vos, B., Verbist, B., Vervoort, A., Poesen, J.,
  Muys, B., Deckers, J., 2009. Spatial structures of soil organic carbon in
  tropical forests-a case study of southeastern tanzania. Catena 77~(1),
  19--27.

\bibitem[{Saby et~al.(2011)Saby, Marchant, Lark, Jolivet, and
  Arrouays}]{saby2011}
Saby, N. P.~A., Marchant, B.~P., Lark, R.~M., Jolivet, C.~C., Arrouays, D.,
  2011. Robust geostatistical prediction of trace elements across france.
  Geoderma 162~(3-4), 303--311.

\bibitem[{Schnebelen et~al.(2004)Schnebelen, Nicoullaud, Bourennane, Couturier,
  Verbeque, Revalier, Bruand, and Ledoux}]{31}
Schnebelen, N., Nicoullaud, B., Bourennane, H., Couturier, A., Verbeque, B.,
  Revalier, C., Bruand, A., Ledoux, E., 2004. The stics model to predict
  nitrate leaching following agricultural practices. Agronomie 24~(6-7),
  423--435.

\bibitem[{Schrumpf et~al.(2011)Schrumpf, Schulze, Kaiser, and
  Schumacher}]{Schrumpf2011}
Schrumpf, M., Schulze, E.~D., Kaiser, K., Schumacher, J., 2011. How accurately
  can soil organic carbon stocks and stock changes be quantified by soil
  inventories? Biogeosciences 8~(5), 1193--1212.

\bibitem[{Spielvogel et~al.(2009)Spielvogel, Prietzel, Auerswald, and
  Koegel-Knabner}]{Spielvogel2009}
Spielvogel, S., Prietzel, J., Auerswald, K., Koegel-Knabner, I., 2009.
  Site-specific spatial patterns of soil organic carbon stocks in different
  landscape units of a high-elevation forest including a site with forest
  dieback. Geoderma 152~(3-4), 218--230.

\bibitem[{Suuster et~al.(2012)Suuster, Ritz, Roostalu, Kolli, and
  Astover}]{Suuster2012}
Suuster, E., Ritz, C., Roostalu, H., Kolli, R., Astover, A., 2012. Modelling
  soil organic carbon concentration of mineral soils in arable land using
  legacy soil data. Eur J Soil Sci 63~(3), 351--359.

\bibitem[{Tomlinson and Milne(2006)}]{Tomlinson2006}
Tomlinson, R.~W., Milne, R.~M., 2006. Soil carbon stocks and land cover in
  northern ireland from 1939 to 2000. Applied Geography 26~(1), 18--39.

\bibitem[{Tornquist et~al.(2009)Tornquist, Giasson, Mielniczuk, Cerri, and
  Bernoux}]{Tornquist2009}
Tornquist, C.~G., Giasson, E., Mielniczuk, J., Cerri, C. E.~P., Bernoux, M.,
  2009. Soil organic carbon stocks of rio grande do sul, brazil. Soil Science
  Society of America Journal 73~(3), 975--982.

\bibitem[{Turner et~al.(1989)Turner, O'Neill, Gardner, and Milne}]{Turner1989}
Turner, M.~G., O'Neill, R.~V., Gardner, R.~H., Milne, B.~T., 1989. Effects of
  changing spatial scale on the analysis of landscape pattern. Landscape
  Ecology 3~(3-4), 153--162.

\bibitem[{UE-SOeS(2006)}]{UE-SOeS2006}
UE-SOeS, 2006. corine land cover. service de l'observation et des statistiques
  (soes) du ministère chargé de l'environnement. Tech. rep.

\bibitem[{UMR-LERFOB and Ifn(2008)}]{pHMapForest2008}
UMR-LERFOB, Ifn, 2008. Carte des ph de surface des sols forestiers. données et
  guide d'utilisation selon le document accord agroparistech (umr lerfob) ifn
  n° 2007-cpa-2-072. Tech. rep.

\bibitem[{Ungar et~al.(2010)Ungar, Staffilani, and Tarocco}]{Ungar2010}
Ungar, F., Staffilani, F., Tarocco, P., 2010. Assessing and mapping topsoil
  organic carbon stock at regional scale: A scorpan kriging approach
  conditional on soil map delineations and land use. Land Degradation and
  Development 21~(6), 565--581.

\bibitem[{van Wesemael et~al.(2010)van Wesemael, Paustian, Meersmans, Goidts,
  Barancikova, and Easter}]{van_Wesemael2010}
van Wesemael, B., Paustian, K., Meersmans, J., Goidts, E., Barancikova, G.,
  Easter, M., 2010. Agricultural management explains historic changes in
  regional soil carbon stocks. Proceedings of the National Academy of Sciences
  of the United States of America 107~(33), 14926--14930.

\bibitem[{Vasques et~al.(2010)Vasques, Grunwald, Comerford, and
  Sickman}]{Vasques2010}
Vasques, G.~M., Grunwald, S., Comerford, N.~B., Sickman, J.~O., 2010. Regional
  modelling of soil carbon at multiple depths within a subtropical watershed.
  Geoderma 156~(3-4), 326--336.

\bibitem[{Webster and Oliver(2007)}]{Webster2007}
Webster, R., Oliver, M.~A., 2007. Geostatistics for Environmental Scientists
  (Statistics in Practice), 2nd Edition. Wiley.

\bibitem[{Wessel and Smith(1991)}]{Wessel1991}
Wessel, P., Smith, W. H.~F., 1991. Free software helps map and display data.
  EOS Tran 16, 441--446.

\bibitem[{Xie et~al.(2011)Xie, Chen, Lei, Yang, Guo, Song, and Zhou}]{Xie2011}
Xie, Y., Chen, T.-b., Lei, M., Yang, J., Guo, Q.-j., Song, B., Zhou, X.-y.,
  2011. Spatial distribution of soil heavy metal pollution estimated by
  different interpolation methods: Accuracy and uncertainty analysis.
  Chemosphere 82~(3), 468--476.

\bibitem[{Xu and Liang(2001)}]{XU2001}
Xu, Q., Liang, Y., 2001. Monte carlo cross-validation. Chemometrics and
  Intelligent laboratory Systems 56~(1), 1--11.

\bibitem[{Yang et~al.(2008)Yang, Fang, Tang, Ji, Zheng, He, and Zhu}]{Yang2008}
Yang, Y.~H., Fang, J.~Y., Tang, Y.~H., Ji, C.~J., Zheng, C.~Y., He, J.~S., Zhu,
  B.~A., 2008. Storage, patterns and controls of soil organic carbon in the
  tibetan grasslands. Global Change Biology 14~(7), 1592--1599.

\bibitem[{Yu et~al.(2007)Yu, Shi, Wang, Sun, Warner, and Liu}]{Yu2007}
Yu, D.~S., Shi, X.~Z., Wang, H.~J., Sun, W.~X., Warner, E.~D., Liu, Q.~H.,
  2007. National scale analysis of soil organic carbon storage in china based
  on chinese soil taxonomy. Pedosphere 17~(1), 11--18.

\bibitem[{Yun-Qiang et~al.(2009)Yun-Qiang, Xing-Chang, Jing-Li, and
  Shun-Ji}]{Wang2009}
Yun-Qiang, W., Xing-Chang, Z., Jing-Li, Z., Shun-Ji, L., 2009. Spatial
  variability of soil organic carbon in a watershed on the loess plateau.
  Pedosphere 19~(4), 486--495.

\bibitem[{Zhao and Shi(2010)}]{Zhao2010}
Zhao, Y.-C., Shi, X.-Z., 2010. Spatial prediction and uncertainty assessment of
  soil organic carbon in hebei province, china. In: Boettinger, J.~L., Howell,
  D.~W., Moore, A.~C. (Eds.), Digital Soil Mapping: Bridging Research,
  Environmental Application, and Operation. Springer Science, Ch.~19, pp.
  227--239.

\end{thebibliography}



%
%


\end{document}